# Entanglement, separability and correlation topology of quantum systems over parametric space of interaction potential

Basudev Nag Chowdhury[1,2*]

[1]Dept. of Electronics & Electrical Communication Engineering, IIT Kharagpur, India

[2]Semiconductor Quantum Device (SQuaD) Lab, QsciT Research, Kolkata, India

*basudevn@gmail.com, bnc@qscit.org

**Abstract**

The standard understanding of formal quantum theory is based upon the belief that the state of two interacting quantum systems can jointly evolve as, either an entangled (such as in case of measurement or decoherence) or a separable (such as in case of single-qubit gate operations) state, through two different processes, *i.e.* "*process 1*" and "*process 2*", respectively, as suggested by von Neumann, although nothing much is known about such processes in terms of physical interaction. The present work, exploring the correlation topologies of two interacting quantum systems in parametric space of their interaction potential, reveals that such "*process 1*" and "*process 2*" are not different kinds of physical interactions but depend on the interaction parameters to result in either an entangled or a separable state. However, under energy conservation restriction, it is impossible to travel from one maximally-entangled state to another in the topological space by continuously varying the interaction parameters without crossing an intermediate separable state, and *vice-versa*. Nevertheless, a maximal entangled state is also shown to be achieved by violating energy conservation utilizing the energy-time uncertainty or a catalytic separable *ancillia*. The work explores the theoretical structures of interaction potential needed to rotate a qubit state on the entire of its Bloch sphere, thereby revealing a novel method of measuring a gate-induced phase to a qubit avoiding the collapse of its state. Further, the interaction leading to local unitary operation on one of the two space-like apart entangled qubits has been illustrated that can manipulate the degree of their non-local entanglement in a controlled manner. Such generalization of "*process 1*" and "*process 2*" in terms of interaction potential to create entanglement/separability suggests a necessary revisit of the fundamental quantum paradoxes and several other quantum limitations including decoherence for advancing the field of quantum technology as a whole.

**Keywords**

Quantum correlation; topology; quantum entanglement; quantum separability; interacting qubits

**Data availability statement**

Data is available on reasonable request from the author.

Quantum entanglement, the most astonishing feature amongst a number of non-classical properties of quantum systems, has been the key resource to the on-going leap toward quantum computation and quantum information technologies [1]. However, despite enormous efforts being made to distinguish between entangled and separable states by experimentally testable methods, through direct or indirect ways, including but not limited to the correlation matrix method [2-3], positive map theory [4-5], entropic inequalities [6], computable cross norm or realignment [7], and numerous other approaches [8-11], the fundamental problem in determining the separability of quantum states still persists. On the other hand, the method of generating steady-state quantum entanglement is hitherto ambiguous due to fundamental energetic limitations originated from correlation-thermodynamic tandem [12-15]. It is indeed leading to widely increasing interest in entanglement engineering from the inter-qubit interaction point of view [15], which further upholds the necessity of investigating entanglement and separability from their rudiments.

At this point it is worthy to mention that, although Einstein, Podolsky, and Rosen (EPR) are usually recognized for exploring the "*spooky*" nature of such quantum entanglement in specially correlated systems [16], von Neumann, prior to the EPR's contribution, described quantum entanglement as an essential component of quantum measurement in general [17]. In fact, following such description the formal quantum theory presumes that, in a measurement-type of interaction, the pure state of any of the subsystems of a composite quantum system (*e.g.* system+apparatus or system+environment) is (in most cases nearly maximally) entangled with its counterpart [17-18]. On the other hand, for other types of interactions such as logic-gate operations of a qubit, the system and the instrument remain in a separable state. Von Neumann distinguishingly termed the former and the latter interactions as '*process 1*' and '*process 2*', respectively [17], although what physically differentiates (and why/how) such processes, *i.e.* from interaction point of view, if any, is still not explicitly understood. Moreover, alongside the hitherto unsolved quantum measurement problem (QMP) associated with the collapse postulate and the non-observance of superposition in classical objects [18], quantum entanglement added to it a more complex problem due to Wigner [19] known as Wigner's friend paradox (WFP). In brief, if Wigner observes a quantum system to be measured by his friend, then the state of the system collapses to one of its basis states along with that of the friend's state, *i.e.* separable, as observed by the friend herself; whereas according to Wigner's perspective, the system and the friend being subsystems of a composite system would be in an entangled state, thereby differing from his friend's observation **[19]**. The notion upholds the inconsistency associated with observer-dependent separability/entanglement. Recently, an extended version of WFP has been proposed for its testability **[20-21]**, and further improved along with its experimental realization **[22-23]**, that has raised fundamental doubt to the objectivity of facts **[24-25]**. Nevertheless, no consensus on such relational objectivity has still been established in this context, while searches for reliable and stable entangled states of quantum systems are continuing for practical quantum computing and quantum teleportation applications **[26-28]**.

In fact, the origin of such problem remains in the conventional understanding of quantum theory which proposes, for instance in a measurement-type of interaction, that if two 2-level systems (*e.g.* two qubits, or, one qubit and two pointers of the measuring apparatus), given by their isolated states (*i.e.* when not interacting) as $|\psi_1\rangle = a|\varphi_1\rangle + b|\varphi_2\rangle$ and $|\psi_2\rangle = c|\chi_1\rangle + d|\chi_2\rangle$, start interacting, then after a certain time of unitary evolution their combined state will be described, for instance, by $|\psi\rangle = e|\varphi_1\rangle \otimes |\chi_1\rangle + f|\varphi_2\rangle \otimes |\chi_2\rangle$ (where $a, b, c, d, e, f$ are normalized complex coefficients) **[17-21, 29]**. We start with questioning that, even if such one-to-one correlation in "*process 1*" is the result of joint evolution, how the systems could get the information that $|\varphi_1\rangle$ would have to correlate to $|\chi_1\rangle$ and not to the state $|\chi_2\rangle$ (and similarly $|\varphi_2\rangle$ with $|\chi_2\rangle$ and not to $|\chi_1\rangle$)? It is imperative to note that the subscripts '1' and '2' are our subjective choices for 'non-invasive labelling' of such states (*i.e.* where no probing is done to selectively label them)! Then, could the final state of the system be different (following our choice of labelling) if we would label them conversely, for instance, $|\psi\rangle$ being different (although symbolically the same, $|\psi\rangle = e|\varphi_1\rangle \otimes |\chi_1\rangle + f|\varphi_2\rangle \otimes |\chi_2\rangle$) while we would have assumed that $|\psi_2\rangle = c|\chi_2\rangle + d|\chi_1\rangle$ instead of $|\psi_2\rangle = c|\chi_1\rangle + d|\chi_2\rangle$? More precisely, the question, in general, is what decides $|\psi\rangle$ to be $|\psi\rangle = e|\varphi_1\rangle \otimes |\chi_1\rangle + f|\varphi_2\rangle \otimes |\chi_2\rangle$ instead of $|\psi\rangle = e|\varphi_1\rangle \otimes |\chi_2\rangle + f|\varphi_2\rangle \otimes |\chi_1\rangle$ or *vice-versa*. This is also not clear why (and most importantly, how) only two (in "*process 1*") of the four possible combinations of basis states (as in "*process 2*") are present in the combined state! (It may be noted that even if the four components are present in general, even then the subsystems of the combined system are assumed to remain in an inseparable state for the former case). It is also interesting to note, further, that the appearance of bonding and anti-bonding states in a diatomic molecule or a coupled-quantum-dot system as well as the formation of energy bands in crystals from its constituent atomic energy levels are obvious outcomes of such inseparability between the subsystems, where even no measurement or decoherence is associated.

However, this is not the case, *e.g.*, for a qubit interacting with a logic-device (the latter being the second system here), that manipulates its basis-state probabilities or phase. For instance, if voltage-gate is used to manipulate the superposition of a double-quantum-dot (DQD) qubit state **[30-34]**, then, in fact, such voltage is applied to the electron in DQD by the positive ions or electrons (for +ve or –ve voltage, respectively) accumulated at the metal/insulator interface. Thus, the states of qubit electron and the metal ion/electron as a combined-state should be entangled in some kind, if it would be similar to the previous case (*i.e.* "*process 1*") such as qubit read-out, which however is not deemed to happen. Otherwise the entire qubit operation would be in question! Then, how can one identify, if possible, whether some particular interaction between systems lead to '*process 1*' or '*process 2*' in von Neumann's terms! Consequently, an apparent question arises about the environmental decoherence too, that why, unlike logic-gates, but rather like the measuring apparatus, the environment must always entangle with the qubit state to decohere it **[35-36]**? All such questions are not only the foundational problems but also extremely important issues to be necessarily addressed in the present context of fast growing quantum computing technology and quantum communication.

In this work, firstly, considering two interacting 2-level quantum systems initially in pure states, we have shown that the final state of the combined-system through their joint unitary evolution (without being coupled to any bath) may become either a separable state or an entangled state (*i.e.* ranging from a minimal degree of entanglement to the maximal one) depending on the very nature and parameters of the interacting potential. The result is then generalized to any numbers of quantum systems with any numbers of quantum states. The nature of interaction potential needed for rotating a qubit state on its Bloch sphere preserving the coherence as well as that for an evolution leading to entanglement has also been illustrated. Subsequently, a 'correlation topology' (based on entanglement/separability) between two 2-level systems has been explored in the parametric space of the interacting potentials, so that such interaction parameters can identify '*process 1*' or '*process 2*' as well as can lead to possible inter-manipulation between entangled or separable quantum states. The energetic constraints for generating entanglement between two qubits from the quantum perspective are also examined without considering coupling with any thermal bath(s). Finally, the method of local manipulation of non-local entanglement of two space-like distant correlated qubits is discussed in the same interaction-based framework.

Let us consider first an example to show on the basis of the nature of interaction potential, how the combined state of two mutually interacting 2-level quantum systems, initially in pure states, can remain in a separable product state through their joint unitary evolution, where unlike **[37]**, such systems are not interacting to any third system such as a 2-qubit logic gate. Suppose the states of two such systems are given by $|\varphi(0)\rangle = 1/\sqrt{2}\left(|\varphi_1\rangle + |\varphi_2\rangle\right)$ and $|\chi(0)\rangle = 1/\sqrt{2}\left(|\chi_1\rangle - |\chi_2\rangle\right)$ (where $\left(|\varphi_1\rangle, |\varphi_2\rangle\right)$ and $\left(|\chi_1\rangle, |\chi_2\rangle\right)$ are orthonormal basis states) just before they start interacting to each other. It may be worthy to note that there is no restriction from the point of view of physical interaction for one of such two interacting systems to be an observer (as in WFP) **[19-21]** or an instrument (*e.g.* a logic gate or a readout device, in general, that may even be a classical object as assumed in the references **[20-23, 29]**). Now, their combined state (prior to interaction) must be given by $|\psi(0)\rangle = 1/2\left(|\varphi_1\rangle + |\varphi_2\rangle\right) \otimes \left(|\chi_1\rangle - |\chi_2\rangle\right)$, associated with their pre-interaction Hamiltonian $\hat{H}_0$, such that $\hat{H}_0|\psi(0)\rangle = [\hat{H}_1 \otimes \hat{I}_2 + \hat{I}_1 \otimes \hat{H}_2]|\psi(0)\rangle = \varepsilon|\psi(0)\rangle$, where $\varepsilon$ is the total energy of the two systems.

Now consider an interaction to start between the systems with a potential of the nature $\hat{V} = V_0\left(|\varphi_1\rangle\langle\varphi_1| \otimes |\chi_1\rangle\langle\chi_1| - |\varphi_2\rangle\langle\varphi_2| \otimes |\chi_2\rangle\langle\chi_2|\right)$, where $V_0$ is real (since the combined system is isolated demanding the potential to be Hermitian). At this point it is imperative to note that the nature of the potential cannot be arbitrary and must respect the energy conservation (*i.e.* $\varepsilon$ is constant) given by another condition: $Tr(\hat{V}\hat{\rho}_{\varphi\chi}(t)) = 0$, where $\hat{\rho}_{\varphi\chi}(t)$ is the density matrix of the combined system at any given instant of time $t$, since there is no third system interacting with the two systems to supply or extract any energy. For instance, $\hat{V} = V_0\left(|\varphi_1\rangle\langle\varphi_1| \otimes |\chi_1\rangle\langle\chi_1| + |\varphi_2\rangle\langle\varphi_2| \otimes |\chi_2\rangle\langle\chi_2|\right)$ is not an allowed interaction potential corresponding to the assumed initial states, owing to $Tr(\hat{V}\hat{\rho}_{\varphi\chi}(0)) \neq 0$, which indicates an extra energy being required from outside the combined system in such case. It may also be interesting to note that such restriction may not be valid for an irreversibly collapsing interaction (as in the

case of an ideal projective measurement) due to the fact that 'time' in such case would not be homogeneous to maintain energy conservation, which, however, to discuss on is beyond the scope of the present letter. Thus, for the considered allowed interaction (which is non-collapsing in general), the combined state after a finite time $\Delta t$, through their joint unitary evolution under the Hamiltonian $\hat{H} = \hat{H}_0 + \hat{V}$, will be given by (see **Appendix-I**),

$$|\psi(\Delta t)\rangle == 1/2\, e^{\left(-\frac{i}{\hbar}\Delta t(\varepsilon+V_0)\right)}\left[\left(|\varphi_1\rangle + e^{\left(\frac{i}{\hbar}\Delta t V_0\right)}|\varphi_2\rangle\right)\otimes\left(|\chi_1\rangle - e^{\left(\frac{i}{\hbar}\Delta t V_0\right)}|\chi_2\rangle\right)\right] \qquad (1)$$

It is evident form Eq. (1) that, even irrespective of the potential-amplitude $V_0$ as well as the time duration of interaction $\Delta t$, the resultant state for such nature of the potential is a product state, *i.e.* separable, and not an entangled state of the two systems. The interacting potential in such case leads to a "*process 2*" that only creates a phase difference between the two basis states of each system, thereby indicating the physical nature of interaction potential needed for coherent manipulation of a qubit to rotate its state through an azimuthal angle on the Bloch sphere. Furthermore, it may be interesting to note that if two qubits can be subjected to interact in such a way, an identical phase difference ($V_0 \Delta t/\hbar$) between the basis states is created to both the qubits maintaining them in pure separable states. It is imperative to note that this is not something like that a same phase gate operation has been performed on both the qubits, since there is no such third party assumed in this work that can be considered as a separate gate. Rather, interestingly the phase introduced to the first qubit is also copied to the second qubit (which itself may be utilized as a phase gate) through their own mutual interaction without any external (*i.e.* third party) intervention. Evidently, the phase of the latter qubit (which is also the phase of first qubit) can be measured without disturbing the state of the first qubit due to their separability, and such process can be repeated for infinitely many times. Thus, this concept can be of potential application in quantum computing where undisturbed phase determination may be of significant interest.

On the other hand, if the interaction potential would instead be of the form $\hat{V} = V_0(i|\varphi_1\rangle\langle\varphi_2| + H.C.)\otimes(|\chi_1\rangle\langle\chi_2| + H.C.)$, the evolved combined-state would be (see **Appendix-I**),

$$\left|\psi(\Delta t)\right\rangle = 1/\sqrt{2}\, e^{\left(-\frac{i}{\hbar}\Delta t\varepsilon\right)}\left[\left(\cos\left(\frac{V_0\Delta t}{\hbar}+\pi/4\right)\left|\varphi_1\right\rangle + \sin\left(\frac{V_0\Delta t}{\hbar}+\pi/4\right)\left|\varphi_2\right\rangle\right)\otimes\left(\left|\chi_1\right\rangle - \left|\chi_2\right\rangle\right)\right] \qquad (2)$$

Eq. (2) shows that such form of potential also leads to a separable state, and not an entangled or mixed state of the two systems, irrespective of the potential-amplitude and interaction duration. However, in this case, the interaction refers to "*process 2*" that can rotate the state of one qubit (*i.e.* the first one here) through a polar angle on the Bloch sphere thereby manipulating the probabilities of its basis states, where the second one remains unaffected and can itself be considered to be the logic gate.

At this point it is worthy to mention that it is apparent that such results (*i.e.* Eq. (1) and (2)) would not change (except the normalization constants) if the initial states were considered to be $\left|\varphi(0)\right\rangle = \left(a_1\left|\varphi_1\right\rangle + a_2\left|\varphi_2\right\rangle\right)$ and $\left|\chi(0)\right\rangle = \left(b_1\left|\chi_1\right\rangle - b_2\left|\chi_2\right\rangle\right)$ in general. Further, if, for instance, the first system would be a 3-level quantum system described by its initial state $\left|\varphi(0)\right\rangle = \left(a_1\left|\varphi_1\right\rangle + a_2\left|\varphi_2\right\rangle + a_3\left|\varphi_3\right\rangle\right)$ in the orthonormal basis $\left(\left|\varphi_1\right\rangle, \left|\varphi_2\right\rangle, \left|\varphi_3\right\rangle\right)$, it is always possible to find another orthonormal basis $\left(\left|\varphi_{12}\right\rangle, \left|\varphi_3\right\rangle\right)$ so that $\left|\varphi(0)\right\rangle = \left(a_{12}\left|\varphi_{12}\right\rangle + a_3\left|\varphi_3\right\rangle\right)$. Thus, an interaction potential of the same forms as above but represented in $\left(\left|\varphi_{12}\right\rangle, \left|\varphi_3\right\rangle, \left\langle\varphi_{12}\right|, \left\langle\varphi_3\right|\right)\otimes\left(\left|\chi_1\right\rangle, \left|\chi_2\right\rangle, \left\langle\chi_1\right|, \left\langle\chi_2\right|\right)$ will result in a similar separable combined-state, indicating that the method can be extended to any numbers of states of the two systems. Further, such two systems can now be assumed as a single joint-system that can interact with a third system through similar interaction potential(s) which will accordingly lead to the relevant separable states. Hence, it is always possible to find interaction potentials so that any numbers of quantum systems with any numbers of quantum states can interact to unitarily evolve in finite time through "*process 2*" as a combined-system with separable states as above.

Nevertheless, such two kinds of interacting potentials illustrated above can be generalized so that they do not lead the interacting qubits to evolving only into the separable states through "*process 2*", which suggests resulting in at least some degree of entanglement between such qubits involving "*process 1*", that may in particular cases be maximal entanglement. For instance, the first kind of interaction, that preserves the basis states of the qubits, generalized as $\hat{V} = V_0\left(\left|\varphi_1\right\rangle\left\langle\varphi_1\right| + \eta\left|\varphi_2\right\rangle\left\langle\varphi_2\right|\right)\otimes\left(\left|\chi_1\right\rangle\left\langle\chi_1\right| + \kappa\left|\chi_2\right\rangle\left\langle\chi_2\right|\right)$ ($V_0, \eta, \kappa$ being real) with its initial

expectation value to be $Tr\left(\hat{V}\hat{\rho}_{\varphi\chi}(0)\right)=\overline{V}(\eta,\kappa)=(1+\eta)(1+\kappa)$, generates the combined state with a concurrence (see **Appendix-II**),

$$C(\eta,\kappa)=(1/2)\left|1-\exp\left(\frac{i}{\hbar}\Delta t V_0(1-\eta)(1-\kappa)\right)\right| \tag{3}$$

Similarly, the second kind of interaction that flips the basis states of interacting qubits can be generalized as $\hat{V}=V_0\left(e^{i\eta}|\varphi_1\rangle\langle\varphi_2|+e^{-i\eta}|\varphi_2\rangle\langle\varphi_1|\right)\otimes\left(e^{i\kappa}|\chi_1\rangle\langle\chi_2|+e^{-i\kappa}|\chi_2\rangle\langle\chi_1|\right)$ ($V_0,\eta,\kappa$ being real) with the expectation value of potential to be $Tr\left(\hat{V}\hat{\rho}_{\varphi\chi}(0)\right)=\overline{V}(\eta,\kappa)=-2[\cos(\eta+\kappa)+\cos(\eta-\kappa)]$ that leads to the concurrence,

$$C(\eta,\kappa)=(1/2)\left|[\cos(\eta+\kappa)-\cos(\eta-\kappa)]\sin\left(\frac{2V_0\Delta t}{\hbar}\right)\right| \tag{4}$$

Such concurrences are plotted in the parametric space $(\eta,\kappa)$ of potential for the first and second kind of interactions in Fig. 1(a) and (b), respectively. The contour lines of initial expectation values of the interacting potential are also shown (the 'black continuous lines') along with that of the zero expectation values (which follow energy conservation) plotted as 'dotted black lines' in such figures. Under such energy conservation, the concurrence value of zero indicating separable states are depicted by 'blue dots' and that of unity referring to maximally entangled states are pointed out by 'red dots'. Thus, we can say that Fig. 1(a) and (b) represent the correlation topology of two qubits in the parametric space of their mutual interaction for the assumed basis-preserving and basis-flipping interactions between the qubits, respectively.

The plots of Fig. 1 show that under the restriction of energy conservation (*i.e.* no external energy is supplied to or extracted from the qubits during to their mutual interaction), the correlation topology for the interaction of first kind (*i.e.* basis-preserving interaction) exhibits only two perpendicular 1D-linear chain lattices of separable and maximally entangled states at alternating topological lattice-cites. On the other hand, such correlation topology for the interaction of second kind (*i.e.* basis-flipping interaction) forms a 2D-lattice with alternating topological lattice-cites for separable and maximally entangled states on the energy-conservation lines. Interestingly, it can be noted that in both cases, one, maintaining energy conservation, cannot continuously vary the interaction parameters $(\eta,\kappa)$ to move from one maximally entangled state to another in the topological space without crossing a separable state, and *vice-versa*. This may be a special feature of correlation topology of two interacting quantum systems. Furthermore, the plots of Fig. 1 upholds that there are no different kinds of particular interactions representing "*process 1*" and "*process 2*", but rather, same form of interactions may lead to "*process 1*" or "*process 2*" depending on the interaction parameters resulting in entangled or separable states, respectively. Such results, therefore, can shed some light on the WFP and its extended versions (especially the experimental results) as well as the fundamentals of decoherence mechanism (and how and whether that can be dealt with), which have presumed that in such cases an unitary evolution would only conclude finally to an entangled or inseparable joint-state, and such view, therefore, needs to be revisited.

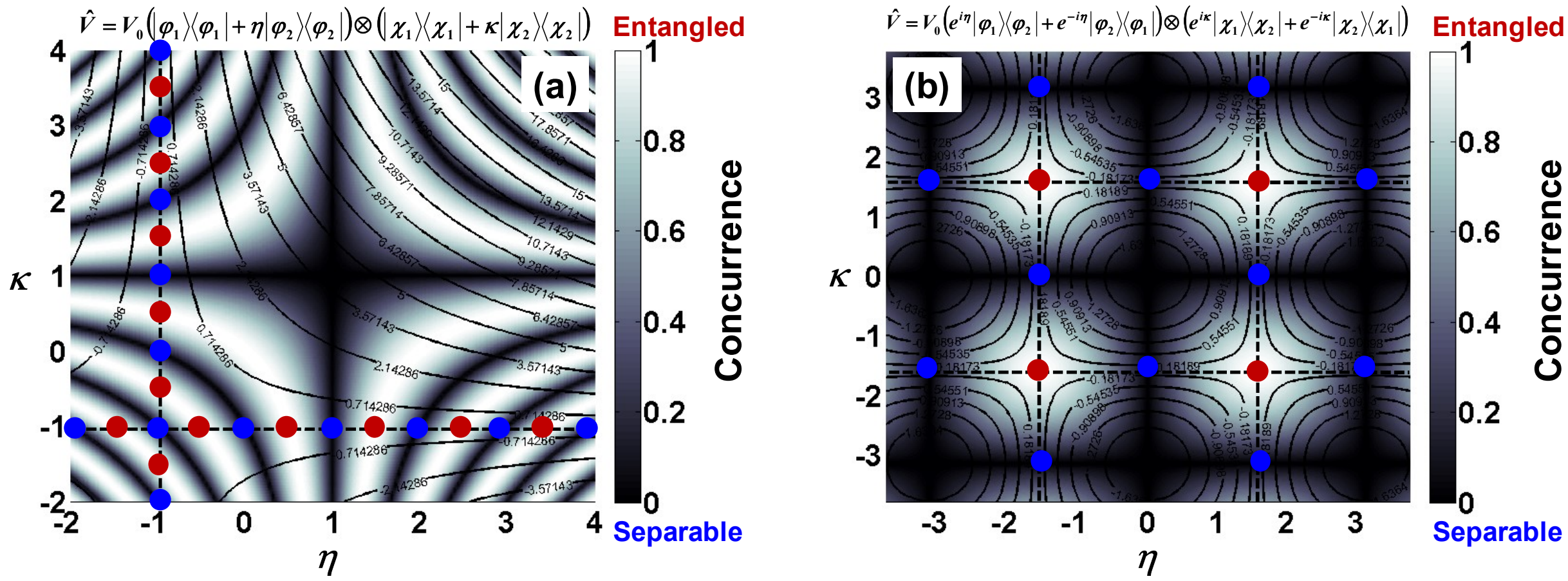

Fig. 1. Correlation topology represented by the plots of concurrence in the potential parameter space $(\eta, \kappa)$, 'black' to 'white' graded colour representing all the possible separable to entangled states accordingly; the 'black' contour lines show the locus of initial expectation values of the interacting potentials in their parameter space, while the 'black' dotted lines indicate the locus of zero expectation values of potentials that are allowed by the restriction of energy conservation; under such energetic limits, the separable and maximally entangled states are represented by 'blue' and 'red' dots, respectively; (a) the first kind of (*i.e.* basis-preserving) interaction; and (b) the second kind of (*i.e.* basis-flipping) interaction.

Further, it can be observed from the plots of Fig. 1(a) and (b) that there are many possible maximally entangled (and also separable) states that can be achieved, intriguingly, by violating the energy conservation, which may be of significant interest from the foundational as well as the applicative quantum information theory. Therefore, to investigate such phenomena, let us consider a different form of a mixed (of first and second kind) interaction, *e.g.* $\hat{V} = \hat{V}_1 + \hat{V}_2 + \hat{V}_3$ such that $\hat{V}_1 = -\varepsilon\left(|\varphi_1\rangle\langle\varphi_1| \otimes |\chi_2\rangle\langle\chi_2| + |\varphi_2\rangle\langle\varphi_2| \otimes |\chi_1\rangle\langle\chi_1|\right)$, $\hat{V}_2 = V_0\left(i|\varphi_1\rangle\langle\varphi_2| \otimes |\chi_1\rangle\langle\chi_1| + H.C.\right)$, and $\hat{V}_3 = V_0\left(i|\varphi_2\rangle\langle\varphi_1| \otimes |\chi_2\rangle\langle\chi_2| + H.C.\right)$, which does not respect $Tr(\hat{V}\hat{\rho}_{\varphi\chi}(0)) = 0$, *i.e.* the energy conservation at initial state (see **Appendix-III**); for instance some energy (positive or negative depending on, for instance, charge or spin of associated particles and interaction) is supplied from outside in order to initiate their mutual interaction. Also suppose that such potential is acting only for a very short duration of time $\delta t \approx \frac{\hbar}{V_0}$ (for instance, $\delta t$ is ~fs if $V_0$ is ~eV). It is imperative to note that such duration is not contradictory to the energy-time uncertainty relation **[38]**; and the combined state after the interaction is obtained to be (see **Appendix-III**),

$$|\psi(\delta t)\rangle = (1/\sqrt{2})\left[|\varphi_1\rangle \otimes |\chi_1\rangle - |\varphi_2\rangle \otimes |\chi_2\rangle\right] \tag{5}$$

which is a pure Bell state, *i.e.* maximally entangled. Such result is consistent with Peres' proposition of short-lasting interaction for achieving an entangled state through interaction (although the present consideration is not restricted to the large interaction potential value compared to the energy of the systems as assumed in Peres' work) **[29]**. However, the most fascinating feature of such nature of the considered interacting potential is that $Tr(\hat{V}\hat{\rho}_{\varphi\chi}(0)) = -\varepsilon/2 \neq 0$, while $Tr(\hat{V}\hat{\rho}_{\varphi\chi}(\delta t)) = 0$. It is imperative to note that such result is not due to any approximation ($\delta t \approx \frac{\hbar}{V_0}$) considered, and rather obvious for the considered form of the interaction potential and the maximally entangled final state as in Eq. (5). Also this is not as apparent as if re-equilibration of the systems after initially switching the potential, since no bath is assumed to couple with the systems and the time-scale of such interaction is considered to be very short. This implies that at the starting instant of interaction, the potential does not satisfy energy conservation, whereas through the combined unitary evolution of two systems (without any bath) it leads to conserve the energy at the end! Our argument on such astonishing phenomenon is that, first, either such interaction can result in a gain/loss of the energy within such small interval of time allowed by the energy-time uncertainty, analogous to virtual propagators at high-energy scattering; or second, in such case physically the two systems must interact with a third system (not a bath), which remains to be pseudo-present to supply and retract such energy during interaction being in a separable sate, *i.e.* a separable quantum *ancillia*, which plays the role analogous to that of a catalyst in a chemical reaction. The latter situation has a close resemblance with the generation of entanglement by coupling with thermal bath(s) **[12-15]**, although no thermodynamic interaction (but only quantum interaction) is considered in the present case. However, we would like to keep it as an open question for future investigations.

Finally, consider a further case, where two such maximally entangled qubits given by their state $|\psi\rangle = (1/\sqrt{2})[|\varphi_1\rangle \otimes |\chi_1\rangle + |\varphi_2\rangle \otimes |\chi_2\rangle]$ are made physically apart to space-like distances, and the second qubit (*i.e.* described by the basis states $|\chi_j\rangle$) is subjected to interact locally with a third qubit (described by $|\eta(0)\rangle = 1/\sqrt{2}\left(|\eta_1\rangle - |\eta_2\rangle\right)$ prior to interaction) in a manner (as in above case-2, see Eq. (2)) such that $\hat{V} = V_0\left(i|\chi_1\rangle\langle\chi_2| + H.C.\right) \otimes \left(|\eta_1\rangle\langle\eta_2| + H.C.\right)$. Then, the evolved combined state of the entangled qubits along with the third qubit will be,

$$|\psi,\eta\rangle \equiv (1/\sqrt{2})\left[\cos\left(\frac{V_0\Delta t}{\hbar} + \pi/4\right)|\varphi_1\rangle \otimes |\chi_1\rangle + \sin\left(\frac{V_0\Delta t}{\hbar} + \pi/4\right)|\varphi_2\rangle \otimes |\chi_2\rangle\right] \otimes \left(|\eta_1\rangle - |\eta_2\rangle\right) \quad (6)$$

Thus, Eq. (6) suggests that in such a way of interacting locally (through "*process 2*") with one of the two space-like apart entangled qubits, it is possible to manipulate their degree of non-local quantum entanglement by such local single-qubit unitary logic operation (considering the third one as a local gate applied on the second qubit only) without destroying the correlation. However, such result is appealing toward further investigation and debate in the context of the previous claims of non-local properties of quantum entanglement to be unaffected by any local manipulations of its subparts **[39]**. The result may be of immense applicative importance for developing novel quantum computing algorithms to achieve superior computing efficiency.

In summary, focusing on the nature of interacting potential between quantum systems, the present work illustrates their possible correlation topologies in the interaction parameter space that generalizes von Neumann's "*process 1*" and "*process 2*" creating entangled and separable states, respectively, revealing that such processes are not fundamentally different but only depend on the interaction parameters. It has explored an interesting feature of such topology that under energy conservation restriction, it is impossible to continuously vary the interaction potential parameters in order to travel from one maximally entangled state to another, bypassing an intermediate separable state in the topological space, and *vice-versa*. It has also been shown that an entangled state can also be achieved by utilizing the energy-time uncertainty or a catalytic separable *ancillia* without coupling with any bath. The work has further investigated the theoretical natures of interaction potential needed to rotating a qubit state on the entire of its Bloch sphere, thereby exploring a novel method of measuring an applied phase introduced to a qubit avoiding the collapse of its state. Notably, the work has illustrated how by locally interacting, through a unitary single-gate operation, with one of the two space-like apart entangled qubits, the degree of their non-local entanglement can be manipulated in a controlled manner without destroying their correlation. The methods can be extended to any number of interacting quantum systems with any number of quantum states in general. In conclusion, the present work fundamentally reveals that by setting up the parameters of interaction potentials of two quantum systems, either "*process 1*" or "*process 2*" can be arranged, and accordingly entanglement or separability can be achieved or manipulated, which can be used to further comprehend and investigate the fundamental paradoxes, such as QMP and WFP (and their updated or extended versions), as well as to address the limitations posed by decoherence.

**Appendix-I**

Suppose initial states of two 2-level systems are,

$$|\varphi(0)\rangle = 1/\sqrt{2}\left(|\varphi_1\rangle + |\varphi_2\rangle\right) \tag{A-I.1}$$

$$|\chi(0)\rangle = 1/\sqrt{2}\left(|\chi_1\rangle - |\chi_2\rangle\right) \tag{A-I.2}$$

where $\left(|\varphi_1\rangle, |\varphi_2\rangle\right)$ and $\left(|\chi_1\rangle, |\chi_2\rangle\right)$ are the complete sets of orthonormal basis states for such systems, respectively, described by the corresponding Hamiltonians given by,

$$\hat{H}_1 = \varepsilon_1\left(|\varphi_1\rangle\langle\varphi_1| + |\varphi_2\rangle\langle\varphi_2|\right) \tag{A-I.3}$$

$$\hat{H}_2 = \varepsilon_2\left(|\chi_1\rangle\langle\chi_1| + |\chi_2\rangle\langle\chi_2|\right) \tag{A-I.4}$$

The combined state of two systems just before they start interacting is apparently,

$$|\psi(0)\rangle = 1/2\left(|\varphi_1\rangle + |\varphi_2\rangle\right) \otimes \left(|\chi_1\rangle - |\chi_2\rangle\right) \tag{A-I.5}$$

with the pre-interaction Hamiltonian given by

$$\hat{H}_0 = [\hat{H}_1 \otimes \hat{I}_2 + \hat{I}_1 \otimes \hat{H}_2] \qquad \text{(A-I.6)}$$

resulting to the total energy $\varepsilon = \varepsilon_1 + \varepsilon_2$. Now, when the systems start to interact, each of them evolves as an open system (which may not be unitary) although the together they evolve unitarily as a combined system. Thus the interaction potential ($\hat{V}$) must have two following properties:

$$\hat{V} = \hat{V}^+ \qquad \text{(A-I.7)}$$

$$Tr(\hat{V}\hat{\rho}_{\varphi\chi}) = 0 \qquad \text{(A-I.8)}$$

where $\hat{\rho}_{\varphi\chi}$ is the density matrix of the combined system (also valid for the initial pre-interaction density matrix), so that during interaction, the Hamiltonian of the combined system remains Hermitian, and the expectation value of interaction potential is zero to conserve the total energy (for non-collapsing interactions). Now we will examine the following cases as a few instances of interaction potentials.

**Case-1:** $\hat{V} = V_0\left(|\varphi_1\rangle\langle\varphi_1| \otimes |\chi_1\rangle\langle\chi_1| - |\varphi_2\rangle\langle\varphi_2| \otimes |\chi_2\rangle\langle\chi_2|\right)$ (A-I.9)

so that

$$\hat{V}\hat{\rho}_{\varphi\chi} \equiv \begin{pmatrix} 1 & 0 & 0 & 0 \\ 0 & 0 & 0 & 0 \\ 0 & 0 & 0 & 0 \\ 0 & 0 & 0 & -1 \end{pmatrix}\begin{pmatrix} 1 & -1 & 1 & -1 \\ -1 & 1 & -1 & 1 \\ 1 & -1 & 1 & -1 \\ -1 & 1 & -1 & 1 \end{pmatrix} = \begin{pmatrix} 1 & -1 & 1 & -1 \\ 0 & 0 & 0 & 0 \\ 0 & 0 & 0 & 0 \\ 1 & -1 & 1 & -1 \end{pmatrix} \qquad \text{(A-I.10)}$$

satisfying Eq. (A-I.8). Now,

$$\hat{V}|\psi(0)\rangle = (1/2)V_0\left(|\varphi_1\rangle \otimes |\chi_1\rangle + |\varphi_2\rangle \otimes |\chi_2\rangle\right)$$

$$\Rightarrow \hat{V}^2|\psi(0)\rangle = (1/2)V_0^{\,2}\left(|\varphi_1\rangle \otimes |\chi_1\rangle - |\varphi_2\rangle \otimes |\chi_2\rangle\right)$$

$$\Rightarrow \hat{V}^3|\psi(0)\rangle = (1/2)V_0^{\,3}\left(|\varphi_1\rangle \otimes |\chi_1\rangle + |\varphi_2\rangle \otimes |\chi_2\rangle\right)$$

… and so on. Thus,

$$|\psi(\Delta t)\rangle = \exp(-\frac{i}{\hbar}\Delta t\hat{H})|\psi(0)\rangle = e^{\left(-\frac{i}{\hbar}\Delta t\varepsilon\right)}\exp(-\frac{i}{\hbar}\Delta t\hat{V})|\psi(0)\rangle$$

$$= e^{\left(-\frac{i}{\hbar}\Delta t\varepsilon\right)}\left[e^{\left(-\frac{i}{\hbar}\Delta t V_0\right)}|\varphi_1\rangle\otimes|\chi_1\rangle-|\varphi_1\rangle\otimes|\chi_2\rangle+|\varphi_2\rangle\otimes|\chi_1\rangle-e^{\left(\frac{i}{\hbar}\Delta t V_0\right)}|\varphi_2\rangle\otimes|\chi_2\rangle\right]$$

$$\Rightarrow|\psi(\Delta t)\rangle=(1/2)e^{\left(-\frac{i}{\hbar}\Delta t(\varepsilon+V_0)\right)}\left[\left(|\varphi_1\rangle+e^{\left(\frac{i}{\hbar}\Delta t V_0\right)}|\varphi_2\rangle\right)\otimes\left(|\chi_1\rangle-e^{\left(\frac{i}{\hbar}\Delta t V_0\right)}|\chi_2\rangle\right)\right] \quad \text{(A-I.11)}$$

**Case-2:** $\hat{V}=V_0\big(i|\varphi_1\rangle\langle\varphi_2|-i|\varphi_2\rangle\langle\varphi_1|\big)\otimes\big(|\chi_1\rangle\langle\chi_2|+|\chi_2\rangle\langle\chi_1|\big)$ (A-I.12)

so that

$$\hat{V}\hat{\rho}_{\varphi\chi}\equiv V_0\begin{pmatrix}0&0&0&i\\0&0&i&0\\0&-i&0&0\\-i&0&0&0\end{pmatrix}\begin{pmatrix}1&-1&1&-1\\-1&1&-1&1\\1&-1&1&-1\\-1&1&-1&1\end{pmatrix}=\begin{pmatrix}-i&\ldots&\ldots&\ldots\\\ldots&-i&\ldots&\ldots\\\ldots&\ldots&i&\ldots\\\ldots&\ldots&\ldots&i\end{pmatrix} \quad \text{(A-I.13)}$$

that also satisfies Eq. (A-I.8). Thus,

$$\hat{V}|\psi(0)\rangle=(1/2)(iV_0)\big(-|\varphi_1\rangle\otimes|\chi_1\rangle+|\varphi_1\rangle\otimes|\chi_2\rangle+|\varphi_2\rangle\otimes|\chi_1\rangle-|\varphi_2\rangle\otimes|\chi_2\rangle\big)$$

$$\Rightarrow\hat{V}^2|\psi(0)\rangle=(1/2)(iV_0)^2\big(-|\varphi_1\rangle\otimes|\chi_1\rangle+|\varphi_1\rangle\otimes|\chi_2\rangle-|\varphi_2\rangle\otimes|\chi_1\rangle+|\varphi_2\rangle\otimes|\chi_2\rangle\big)$$

$$\Rightarrow\hat{V}^3|\psi(0)\rangle=(1/2)(iV_0)^3\big(|\varphi_1\rangle\otimes|\chi_1\rangle-|\varphi_1\rangle\otimes|\chi_2\rangle-|\varphi_2\rangle\otimes|\chi_1\rangle+|\varphi_2\rangle\otimes|\chi_2\rangle\big)$$

… and so on. Considering $x=\left(-\frac{i}{\hbar}\Delta t\right)(iV_0)=\left(\frac{V_0\Delta t}{\hbar}\right)$, the coefficient of the above four terms in the expression of $\exp\left(-\frac{i}{\hbar}\Delta t\hat{V}\right)|\psi(0)\rangle$ are,

$$|\varphi_1\rangle\otimes|\chi_1\rangle\rightarrow\left(1-x-\frac{x^2}{2!}+\frac{x^3}{3!}+\frac{x^4}{4!}-\ldots\right)=\sqrt{2}\cos(x+\pi/4)$$

$$|\varphi_1\rangle\otimes|\chi_2\rangle\rightarrow-\left(1-x-\frac{x^2}{2!}+\frac{x^3}{3!}+\frac{x^4}{4!}-\ldots\right)=-\sqrt{2}\cos(x+\pi/4)$$

$$|\varphi_2\rangle\otimes|\chi_1\rangle\rightarrow\left(1+x-\frac{x^2}{2!}-\frac{x^3}{3!}+\frac{x^4}{4!}+\ldots\right)=\sqrt{2}\sin(x+\pi/4)$$

$$|\varphi_2\rangle \otimes |\chi_1\rangle \to -\left(1 + x - \frac{x^2}{2!} - \frac{x^3}{3!} + \frac{x^4}{4!} + \ldots\right) = -\sqrt{2}\sin(x + \pi/4)$$

Thus,

$$|\psi(\Delta t)\rangle = \frac{1}{\sqrt{2}} e^{\left(-\frac{i}{\hbar}\Delta t\varepsilon\right)} \left[ \left( \cos\left(\frac{V_0 \Delta t}{\hbar} + \pi/4\right) |\varphi_1\rangle + \sin\left(\frac{V_0 \Delta t}{\hbar} + \pi/4\right) |\varphi_2\rangle \right) \otimes \left( |\chi_1\rangle - |\chi_2\rangle \right) \right] \quad \text{(A-I.14)}$$

**Appendix-II**

**General case-1:** The interacting potential is given by,

$$\hat{V} = V_0 \left( |\varphi_1\rangle\langle\varphi_1| + \eta |\varphi_2\rangle\langle\varphi_2| \right) \otimes \left( |\chi_1\rangle\langle\chi_1| + \kappa |\chi_2\rangle\langle\chi_2| \right) \quad \text{(A-II.1)}$$

where $V_0, \eta, \kappa$ are real leading to,

$$Tr\left(\hat{V}\hat{\rho}_{\varphi\chi}(0)\right) = (1+\eta)(1+\kappa) \quad \text{(A-II.2)}$$

Now,

$$\hat{V}|\psi(0)\rangle = (1/2)V_0 \left( |\varphi_1\rangle + \eta |\varphi_2\rangle \right) \otimes \left( |\chi_1\rangle - \kappa |\chi_2\rangle \right)$$

$$\Rightarrow \hat{V}^2 |\psi(0)\rangle = (1/2)V_0^2 \left( |\varphi_1\rangle + \eta^2 |\varphi_2\rangle \right) \otimes \left( |\chi_1\rangle - \kappa^2 |\chi_2\rangle \right)$$

…

Thus, the combined evolved state at time $\Delta t$ is obtained to be,

$$|\psi(\Delta t)\rangle = (1/2) e^{\left(-\frac{i}{\hbar}\Delta t\varepsilon\right)} \left[ \exp\left(-\frac{i}{\hbar}\Delta t V_0\right) |\varphi_1\rangle \otimes |\chi_1\rangle - \exp\left(-\frac{i}{\hbar}\Delta t V_0 \kappa\right) |\varphi_1\rangle \otimes |\chi_2\rangle \right.$$

$$\left. + \exp\left(-\frac{i}{\hbar}\Delta t V_0 \eta\right) |\varphi_2\rangle \otimes |\chi_1\rangle - \exp\left(-\frac{i}{\hbar}\Delta t V_0 \eta\kappa\right) |\varphi_2\rangle \otimes |\chi_2\rangle \right] \quad \text{(A-II.3)}$$

Hence, the concurrence is obtained to be,

$$C = 2\left\| \left[ -(1/4)\exp\left(-\frac{i}{\hbar}\Delta t V_0 (1+\eta\kappa)\right) + (1/4)\exp\left(-\frac{i}{\hbar}\Delta t V_0 (\eta+\kappa)\right) \right] \right\|$$

$$\Rightarrow C = (1/2)\left\| \left[ 1 - \exp\left(\frac{i}{\hbar}\Delta t V_0 (1-\eta)(1-\kappa)\right) \right] \right\| \quad \text{(A-II.4)}$$

**General case-2:** The interacting potential is given by,

$$\hat{V} = V_0\left(e^{i\eta}|\varphi_1\rangle\langle\varphi_2| + e^{-i\eta}|\varphi_2\rangle\langle\varphi_1|\right)\otimes\left(e^{i\kappa}|\chi_1\rangle\langle\chi_2| + e^{-i\kappa}|\chi_2\rangle\langle\chi_1|\right) \qquad \text{(A-II.5)}$$

where $V_0, \eta, \kappa$ are real leading to,

$$Tr\left(\hat{V}\hat{\rho}_{\varphi\chi}(0)\right) = -2[\cos(\eta+\kappa) + \cos(\eta-\kappa)] \qquad \text{(A-II.6)}$$

$$\Rightarrow \hat{V}|\psi(0)\rangle = (1/2)V_0\left(-e^{i(\eta+\kappa)}|\varphi_1\rangle\otimes|\chi_1\rangle + e^{i(\eta-\kappa)}|\varphi_1\rangle\otimes|\chi_2\rangle - e^{-i(\eta-\kappa)}|\varphi_2\rangle\otimes|\chi_1\rangle + e^{-i(\eta+\kappa)}|\varphi_2\rangle\otimes|\chi_2\rangle\right)$$

$$\Rightarrow \hat{V}^2|\psi(0)\rangle = (1/2)V_0^{\,2}\left(|\varphi_1\rangle\otimes|\chi_1\rangle - |\varphi_1\rangle\otimes|\chi_2\rangle + |\varphi_2\rangle\otimes|\chi_1\rangle - |\varphi_2\rangle\otimes|\chi_2\rangle\right)$$

…

$$\begin{aligned}|\psi(\Delta t)\rangle = (1/2)e^{\left(-\frac{i}{\hbar}\Delta t\varepsilon\right)}\Bigg[&\left(\cos\left(\frac{V_0\Delta t}{\hbar}\right) - ie^{i(\eta+\kappa)}\sin\left(\frac{V_0\Delta t}{\hbar}\right)\right)|\varphi_1\rangle\otimes|\chi_1\rangle \\ &-\left(\cos\left(\frac{V_0\Delta t}{\hbar}\right) - ie^{i(\eta-\kappa)}\sin\left(\frac{V_0\Delta t}{\hbar}\right)\right)|\varphi_1\rangle\otimes|\chi_2\rangle \\ &+\left(\cos\left(\frac{V_0\Delta t}{\hbar}\right) - ie^{-i(\eta-\kappa)}\sin\left(\frac{V_0\Delta t}{\hbar}\right)\right)|\varphi_2\rangle\otimes|\chi_1\rangle \\ &-\left(\cos\left(\frac{V_0\Delta t}{\hbar}\right) - ie^{-i(\eta+\kappa)}\sin\left(\frac{V_0\Delta t}{\hbar}\right)\right)|\varphi_2\rangle\otimes|\chi_2\rangle\Bigg]\end{aligned}$$

Thus the concurrence is obtained to be,

$$C = 2\left|-(1/4)\left[\cos\left(\frac{2V_0\Delta t}{\hbar}\right) - i\cos(\eta+\kappa)\sin\left(\frac{2V_0\Delta t}{\hbar}\right)\right] + (1/4)\left[\cos\left(\frac{2V_0\Delta t}{\hbar}\right) - i\cos(\eta-\kappa)\sin\left(\frac{2V_0\Delta t}{\hbar}\right)\right]\right|$$

$$\Rightarrow C = (1/2)\left|[\cos(\eta+\kappa) - \cos(\eta-\kappa)]\sin\left(\frac{2V_0\Delta t}{\hbar}\right)\right| \qquad \text{(A-II.7)}$$

**Appendix-II**

Suppose the interaction potential, which acts for a very short period of time $\delta t \approx \dfrac{\hbar}{V_0}$, is given by,

$$\hat{V} = -\varepsilon\left(|\varphi_1\rangle\langle\varphi_1|\otimes|\chi_2\rangle\langle\chi_2| + |\varphi_2\rangle\langle\varphi_2|\otimes|\chi_1\rangle\langle\chi_1|\right)$$

$$+V_0\left(i|\varphi_1\rangle\langle\varphi_2|\otimes|\chi_1\rangle\langle\chi_1| + H.C.\right) + V_0\left(i|\varphi_2\rangle\langle\varphi_1|\otimes|\chi_2\rangle\langle\chi_2| + H.C.\right) \qquad \text{(A-III.1)}$$

so that,

$$\hat{V}\hat{\rho}_{\varphi\chi} \equiv \begin{pmatrix} 0 & 0 & iV_0 & 0 \\ 0 & -\varepsilon & 0 & -iV_0 \\ -iV_0 & 0 & -\varepsilon & 0 \\ 0 & iV_0 & 0 & 0 \end{pmatrix} \begin{pmatrix} 1 & -1 & 1 & -1 \\ -1 & 1 & -1 & 1 \\ 1 & -1 & 1 & -1 \\ -1 & 1 & -1 & 1 \end{pmatrix} = \begin{pmatrix} iV_0 & \dots & \dots & \dots \\ \dots & -\varepsilon - iV_0 & \dots & \dots \\ \dots & \dots & -\varepsilon - iV_0 & \dots \\ \dots & \dots & \dots & iV_0 \end{pmatrix}$$

leading to

$$Tr(\hat{V}\hat{\rho}_{\varphi\chi}) = -\varepsilon/2 \qquad \text{(A-III.2)}$$

which does not satisfies Eq. (A-I.8). Now,

$$\hat{V}|\psi(0)\rangle = (1/2)\left(iV_0|\varphi_1\rangle\otimes|\chi_1\rangle + (\varepsilon + iV_0)\left(|\varphi_1\rangle\otimes|\chi_2\rangle - |\varphi_2\rangle\otimes|\chi_1\rangle\right) - iV_0|\varphi_2\rangle\otimes|\chi_2\rangle\right) \quad \text{(A-III.3)}$$

The combined state just after the interaction during the time $\delta t \approx \dfrac{\hbar}{V_0}$ is achieved to be,

$$|\psi(\delta t)\rangle = \left[1 - \frac{i}{\hbar}\delta t\hat{H}\right]|\psi(0)\rangle = \left[1 - \frac{i}{\hbar}\delta t\hat{H}_0\right]|\psi(0)\rangle + \left[-\frac{i}{\hbar}\delta t\hat{V}\right]|\psi(0)\rangle$$

$$= (1/2)\left[1 - \frac{i}{\hbar}\delta t\varepsilon\right]\left[|\varphi_1\rangle\otimes|\chi_1\rangle - \left(|\varphi_1\rangle\otimes|\chi_2\rangle - |\varphi_2\rangle\otimes|\chi_1\rangle\right) - |\varphi_2\rangle\otimes|\chi_2\rangle\right]$$

$$+ (1/2)\left[-\frac{i}{\hbar}\delta t\right]\left[iV_0|\varphi_1\rangle\otimes|\chi_1\rangle + (\varepsilon + iV_0)\left(|\varphi_1\rangle\otimes|\chi_2\rangle - |\varphi_2\rangle\otimes|\chi_1\rangle\right) - iV_0|\varphi_2\rangle\otimes|\chi_2\rangle\right]$$

$$= (1/2)\left[1 - \frac{i}{\hbar}\delta t\varepsilon\right]\left[|\varphi_1\rangle\otimes|\chi_1\rangle - \left(|\varphi_1\rangle\otimes|\chi_2\rangle - |\varphi_2\rangle\otimes|\chi_1\rangle\right) - |\varphi_2\rangle\otimes|\chi_2\rangle\right]$$

$$+ (1/2)\left[|\varphi_1\rangle\otimes|\chi_1\rangle + \left[1 - \frac{i}{\hbar}\delta t\varepsilon\right]\left(|\varphi_1\rangle\otimes|\chi_2\rangle - |\varphi_2\rangle\otimes|\chi_1\rangle\right) - |\varphi_2\rangle\otimes|\chi_2\rangle\right] \quad \left[\text{since } \delta t \approx \frac{\hbar}{V_0}\right]$$

$$= \left[1-(1/2)\frac{i}{\hbar}\delta t\varepsilon\right]\left[|\varphi_1\rangle\otimes|\chi_1\rangle-|\varphi_2\rangle\otimes|\chi_2\rangle\right]$$

which after normalization gives rise to,

$$|\psi(\delta t)\rangle = (1/\sqrt{2})\left[|\varphi_1\rangle\otimes|\chi_1\rangle-|\varphi_2\rangle\otimes|\chi_2\rangle\right] \qquad \text{(A-III.4)}$$

Interestingly,

$$\hat{V}\hat{\rho}_{\varphi\chi}(\delta t) \equiv \begin{pmatrix} 0 & 0 & iV_0 & 0 \\ 0 & -\varepsilon & 0 & -iV_0 \\ -iV_0 & 0 & -\varepsilon & 0 \\ 0 & iV_0 & 0 & 0 \end{pmatrix}\begin{pmatrix} 1 & 0 & 0 & -1 \\ 0 & 0 & 0 & 0 \\ 0 & 0 & 0 & 0 \\ -1 & 0 & 0 & 1 \end{pmatrix} = \begin{pmatrix} 0 & \ldots & \ldots & \ldots \\ \ldots & 0 & \ldots & \ldots \\ \ldots & \ldots & 0 & \ldots \\ \ldots & \ldots & \ldots & 0 \end{pmatrix}$$

Resulting in,

$$Tr\left(\hat{V}\hat{\rho}_{\varphi\chi}(\delta t)\right) = 0 \qquad \text{(A-III.5)}$$

**Acknowledgement**

The author acknowledges ICMR, India for providing Principal Research Scientist fellowship through IIT, Kharagpur, India, and is grateful to Basudev Lahiri for providing relevant infrastructure to conduct the research work.

**Conflict of interest**

The author has no conflicts to disclose

**Funding declaration**

The author declares that the present research received no specific grant from any funding agency in the public, commercial, or not-for-profit sectors.

**References**

1. Lu, Yang, Alexander Sigov, Leonid Ratkin, Leonid A. Ivanov, and Min Zuo. "Quantum computing and industrial information integration: A review." *Journal of Industrial Information Integration* (2023): 100511.
2. De Vicente, Julio Í. "Separability criteria based on the Bloch representation of density matrices." (2007).

3. De Vicente, Julio I. "Further results on entanglement detection and quantification from the correlation matrix criterion." *Journal of Physics A: Mathematical and Theoretical* 41, no. 6 (2008): 065309.

4. Peres, Asher. "Separability criterion for density matrices." *Physical Review Letters* 77, no. 8 (1996): 1413.

5. Horodecki, Michal, Pawel Horodecki, and Ryszard Horodecki. "On the necessary and sufficient conditions for separability of mixed quantum states." *Phys. Lett. A* 223, no. 1 (1996).

6. Coles, Patrick J., Mario Berta, Marco Tomamichel, and Stephanie Wehner. "Entropic uncertainty relations and their applications." *Reviews of Modern Physics* 89, no. 1 (2017): 015002.

7. Zhang, Cheng-Jie, Yong-Sheng Zhang, Shun Zhang, and Guang-Can Guo. "Entanglement detection beyond the computable cross-norm or realignment criterion." *Physical Review A—Atomic, Molecular, and Optical Physics* 77, no. 6 (2008): 060301.

8. Horodecki, Ryszard, Paweł Horodecki, Michał Horodecki, and Karol Horodecki. "Quantum entanglement." *Reviews of modern physics* 81, no. 2 (2009): 865-942.

9. Gühne, Otfried, and Géza Tóth. "Entanglement detection." *Physics Reports* 474, no. 1-6 (2009): 1-75.

10. Ghosh, Roopayan, and Sougato Bose. "Separability criterion using one observable for special states: Entanglement detection via quantum quench." *Physical Review Research* 6, no. 2 (2024): 023132.

11. Ureña, Julio, Antonio Sojo, Juani Bermejo-Vega, and Daniel Manzano. "Entanglement detection with classical deep neural networks." *Scientific Reports* 14, no. 1 (2024): 18109.

12. Brask, Jonatan Bohr, Géraldine Haack, Nicolas Brunner, and Marcus Huber. "Autonomous quantum thermal machine for generating steady-state entanglement." *New Journal of Physics* 17, no. 11 (2015): 113029.

13. Friis, Nicolai, Marcus Huber, and Martí Perarnau-Llobet. "Energetics of correlations in interacting systems." *Physical Review E* 93, no. 4 (2016): 042135.

14. Khandelwal, Shishir, Björn Annby-Andersson, Giovanni Francesco Diotallevi, Andreas Wacker, and Armin Tavakoli. "Maximal steady-state entanglement in autonomous quantum thermal machines." *arXiv preprint arXiv:2401.01776* (2024).

15. De Oliveira Junior, A., Jeongrak Son, Jakub Czartowski, and Nelly HY Ng. "Entanglement generation from athermality." *Physical Review Research* 6, no. 3 (2024): 033236.

16. Einstein, Albert, Boris Podolsky, and Nathan Rosen. "Can quantum-mechanical description of physical reality be considered complete?." *Physical review* 47, no. 10 (1935): 777.

17. Von Neumann, John. *Mathematical foundations of quantum mechanics: New edition*. Vol. 53. Princeton university press, 2018.

18. Schrödinger, Erwin. "The present status of quantum mechanics." *Die Naturwissenschaften* 23, no. 48 (1935): 1-26.

19. Wigner, Eugene P. "Remarks on the mind-body question." In *Philosophical reflections and syntheses*, pp. 247-260. Berlin, Heidelberg: Springer Berlin Heidelberg, 1995.

20. Brukner, Časlav. "On the quantum measurement problem." *Quantum [un] speakables II: half a century of Bell's theorem* (2017): 95-117.

21. Brukner, Časlav. "A no-go theorem for observer-independent facts." *Entropy* 20, no. 5 (2018): 350.

22. Proietti, Massimiliano, Alexander Pickston, Francesco Graffitti, Peter Barrow, Dmytro Kundys, Cyril Branciard, Martin Ringbauer, and Alessandro Fedrizzi. "Experimental test of local observer independence." *Science advances* 5, no. 9 (2019): eaaw9832.

23. Bong, Kok-Wei, Aníbal Utreras-Alarcón, Farzad Ghafari, Yeong-Cherng Liang, Nora Tischler, Eric G. Cavalcanti, Geoff J. Pryde, and Howard M. Wiseman. "A strong no-go theorem on the Wigner's friend paradox." *Nature Physics* 16, no. 12 (2020): 1199-1205.

24. Brukner, Časlav. "Facts are relative." *Nature Physics* 16, no. 12 (2020): 1172-1174.

25. Brukner, Časlav. "Wigner's friend and relational objectivity." *Nature Reviews Physics* 4, no. 10 (2022): 628-630.

26. Karamlou, Amir H., Ilan T. Rosen, Sarah E. Muschinske, Cora N. Barrett, Agustin Di Paolo, Leon Ding, Patrick M. Harrington et al. "Probing entanglement in a 2D hard-core Bose–Hubbard lattice." *Nature* (2024): 1-6.

27. Chin, Seungbeom, Yong-Su Kim, and Marcin Karczewski. "Shortcut to multipartite entanglement generation: A graph approach to boson subtractions." *npj Quantum Information* 10, no. 1 (2024): 67.

28. Kam, John F., Haiyue Kang, Charles D. Hill, Gary J. Mooney, and Lloyd CL Hollenberg. "Characterization of entanglement on superconducting quantum computers of up to 414 qubits." *Physical Review Research* 6, no. 3 (2024): 033155.

29. Peres, Asher. "Can we undo quantum measurements?." *Physical Review D* 22, no. 4 (1980): 879.

30. Gorman, J., D. G. Hasko, and D. A. Williams. "Charge-qubit operation of an isolated double quantum dot." *Physical review letters* 95, no. 9 (2005): 090502.

31. Shi, Zhan, C. B. Simmons, J. R. Prance, John King Gamble, Teck Seng Koh, Yun-Pil Shim, Xuedong Hu et al. "Fast hybrid silicon double-quantum-dot qubit." *Physical review letters* 108, no. 14 (2012): 140503.

32. Kim, Dohun, Zhan Shi, C. B. Simmons, D. R. Ward, J. R. Prance, Teck Seng Koh, John King Gamble et al. "Quantum control and process tomography of a semiconductor quantum dot hybrid qubit." *Nature* 511, no. 7507 (2014): 70-74.

33. Nag Chowdhury, Basudev, and Sanatan Chattopadhyay. "Dual-Gate GaAs-Nanowire FET for Room Temperature Charge-Qubit Operation: A NEGF Approach." *Advanced Quantum Technologies* 6, no. 4 (2023): 2200072.

34. Steffensen, Gorm Ole, and A. Levy Yeyati. "Yu-Shiba-Rusinov–Bond Qubit in a Double Quantum Dot with Circuit-QED Operation." *PRX Quantum* 6, no. 2 (2025): 020329.

35. Zurek, Wojciech Hubert. "Decoherence, einselection, and the quantum origins of the classical." *Reviews of modern physics* 75, no. 3 (2003): 715.

36. Schlosshauer, Maximilian. "Quantum decoherence." *Physics Reports* 831 (2019): 1-57.

37. Kraus, Barbara, and J. Ignacio Cirac. "Optimal creation of entanglement using a two-qubit gate." *Physical Review A* 63, no. 6 (2001): 062309.

38. Aharonov, Yakir, and David Bohm. "Time in the quantum theory and the uncertainty relation for time and energy." *Physical Review* 122, no. 5 (1961): 1649.

39. Makhlin, Yuriy. "Nonlocal properties of two-qubit gates and mixed states, and the optimization of quantum computations." *Quantum Information Processing* 1, no. 4 (2002): 243-252.